\documentclass[a4paper]{article}
\usepackage[margin=25mm]{geometry}
\usepackage{url}
\usepackage{import}
\usepackage{amsfonts,amssymb,amsmath,amsthm,dsfont}
\usepackage{xcolor}
\usepackage{comment}
\usepackage{algorithmic}
\usepackage{booktabs}
\usepackage{bm}
\usepackage[normalem]{ulem}
\usepackage{multirow}
\usepackage{textcomp}
\usepackage{adjustbox}
\usepackage{mathtools}

\providecommand{\keywords}[1]
{
  \small
  \textbf{\textit{Keywords---}} #1
}
\begin{document}

\title{Leveraging the structure of musical preference in\\ content-aware music recommendation}
\date{}
\author{Paul~Magron\thanks{IRIT, Université de Toulouse, CNRS, Toulouse, France (e-mail: firstname.lastname@irit.fr).} ,
        Cédric~Févotte\footnotemark[1]
}

\maketitle
\begin{abstract}
State-of-the-art music recommendation systems are based on collaborative filtering, which predicts a user's interest from his listening habits and similarities with other users' profiles. These approaches are agnostic to the song content, and therefore face the cold-start problem: they cannot recommend novel songs without listening history. To tackle this issue, content-aware recommendation incorporates information about the songs that can be used for recommending new items. Most methods falling in this category exploit either user-annotated tags, acoustic features or deeply-learned features. Consequently, these content features do not have a clear musical meaning, thus they are not necessarily relevant from a musical preference perspective. In this work, we propose instead to leverage a model of musical preference which originates from the field of music psychology. From low-level acoustic features we extract three factors (arousal, valence and depth), which have been shown appropriate for describing musical taste. Then we integrate those into a collaborative filtering framework for content-aware music recommendation. Experiments conducted on large-scale data show that this approach is able to address the cold-start problem, while using a compact and meaningful set of musical features.
\end{abstract}
\section{Introduction}

Music recommendation~\cite{Schedl2015} consists in predicting users' listening habits in order to suggest them novel tracks that they might enjoy. This task, which is at the core of many commercial platforms, has been extensively investigated, but remains challenging due to the complexity of music and to the lack of explicit and reliable users feedback~\cite{Hu2008}.
Among these many challenges, accounting for contextual information is of paramount importance, so that recommendations are adequate with a specific situation or location~\cite{Gillhofer2015}. Besides, music recommender systems face the \emph{cold-start} problem~\cite{Schein2002}: when a new song is added to a music catalog, there is no interaction data between the users and this song. Consequently, the system cannot properly recommend this new item to existing users. This problem, which is still considered as a major challenge in music recommendation research~\cite{Schedl2018}, is the topic of interest of the present paper.

State-of-the-art approaches for music recommendation are based on collaborative filtering~\cite{Hu2008}, a family of techniques which rely solely on users' listening history: the interest of a given user for a given song is predicted using similarities between various user profiles. The users' feedback are most often implicit and in the form of \emph{playcounts}, that is, how many times a given user has listened to a particular song. This listening history is however noisy and lacks negative feedback data~\cite{Hu2008}. To alleviate this problem, weighted matrix factorization (WMF) techniques~\cite{Mnih2007,Koren2009} process binarized data computed from the raw playcounts, and associate a measure of \emph{confidence} to these binarized feedback. This family of techniques has shown good performance in music recommender systems~\cite{Liang2015}.
More recently, with the advent of deep learning, collaborative filtering techniques using deep neural networks (DNNs) have been proposed and yield promising results~\cite{He2017,Chen2019}. However, WMF techniques remain a solid choice, as they are light, flexible, and still compete with neural collaborative filtering approaches~\cite{Dacrema2019}.

Such approaches are however agnostic to any form of song-related content, which becomes a major issue for new items: a collaborative filtering method is not able to recommend a song without listening history, and therefore face the cold-start problem. This problem has been tackled with content-based methods, which aim to exploit additional information about the items for recommendation~\cite{Yoshii2006,Chong2011,Gopalan2014}. For instance,~\cite{Gouvert2018} exploits tags in a co-factorization framework, deep content-based approaches~\cite{Oord2013,Wang2014} use acoustic features in conjunction with collaborative filtering, and~\cite{Liang2015} uses the last hidden layer of an auto-tagging DNN as content feature.

Even though these approaches have been shown useful for addressing the cold-start problem, these content features lack a clear musical meaning. Consequently, they are not necessarily relevant in terms of musical preference, and appropriate for recommendation. On the other hand, research has been conducted on the structure of musical preference~\cite{Rentfrow2011}. Recent studies~\cite{Greenberg2016,Fricke2018,Fricke2019} notably show that musical preference can be described by using a set of three factors termed \emph{arousal}, \emph{valence} and \emph{depth} (AVD). The usefulness of musical preference models for recommendation has been pointed out in~\cite{Laplante2014}, but to the best of our knowledge, it has only been exploited in~\cite{Soleymani2015}. However, this approach is not based on collaborative filtering and relies on expert ratings for estimating a genre-specific musical preference model, which hinders its deployment at a larger scale.

In this paper, we propose to leverage the AVD model of musical preference in music recommendation based on collaborative filtering. Unlike prior work using tag-based or deeply-learned features, we claim that using descriptors that characterize musical preference is appropriate for recommendation. We therefore aim at bridging the gap between music recommendation and music psychology, which has notably been motivated in a recent survey~\cite{Schedl2018}. We compute the AVD factors and we integrate them as content features into a WMF-based framework for collaborative filtering. We experimentally assess the potential of this method for cold-start recommendation, where it notably outperforms a pure content-based method.

The rest of this paper is structured as follows. Section~\ref{sec:model} presents the collaborative filtering framework. Section~\ref{sec:content} describes the AVD model. Section~\ref{sec:exp} details the music recommendation experiments. Finally, Section~\ref{sec:conclusion} draws some concluding remarks.

\section{Collaborative filtering model}
\label{sec:model}

In this section, we present the content-aware collaborative filtering framework~\cite{Liang2015} on which our method is based. Even though the literature on matrix factorization models is quite abundant, we adopted this WMF framework as it was shown appropriate for handling implicit feedback data~\cite{Hu2008,Oord2013}. We first describe the baseline WMF model, and then present its content-aware extension.

\subsection{Weighted matrix factorization}

We consider playcount data $\mathbf{Y} \in \mathbb{R}^{U \times I}$, where $U$ and $I$ respectively denote the number of users and items. The $(u,i)$-th entry of $\mathbf{Y}$, denoted $y_{u,i}$, is the number of times the song $i$ has been listened to by user $u$.

WMF~\cite{Hu2008,Koren2009} models the data as the product of two low-dimensional factors: a (transposed) user preferences matrix $\mathbf{W} \in \mathbb{R}^{K \times U}$ and an item attributes matrix $\mathbf{H} \in \mathbb{R}^{K \times I}$. The rank $K$ of the decomposition is chosen such that $K(U+I) << UI$ to ensure dimensionality reduction. In a probabilistic framework~\cite{Mnih2007}, these factors are usually assumed to be drawn from centered normal distributions with scaling parameters $\lambda_W$ and $\lambda_H$:
\begin{equation}
\begin{aligned}
     \forall u \in \{1,...,U \} &\text{, } \mathbf{w}_{u} \sim \mathcal{N}( 0,  \lambda_W ^{-1} \mathbf{I}_K ), \\
     \forall i \in \{1,...,I \} &\text{, } \mathbf{h}_{i} \sim \mathcal{N}( 0,  \lambda_H ^{-1} \mathbf{I}_K ),
\end{aligned}
\label{eq:prior_w_h}
\end{equation}
where $\mathbf{w}_{u}$ (resp. $\mathbf{h}_{i}$) is the $u$-th (resp. $i$-th) column of $\mathbf{W}$ (resp. $\mathbf{H}$), and $\mathbf{I}_K$ is the identity matrix of dimension $K$. In order to better account for the over-dispersed nature of the raw data, it is common to consider the binarized playcount matrix $\mathbf{R}$, which indicates whether a user has listened to a song more than a certain amount of times or not~\cite{Liang2015}. Note that alternative strategies have been proposed to address this issue, notably by crafting refined statistical models based on the Poisson or compound Poisson distributions~\cite{Gopalan2015,Basbug2016,Gouvert2019}.

\noindent The generative process for the observed data is then:
\begin{equation}
    \forall u \text{, } i \text{, } r_{u,i} \sim \mathcal{N}( \mathbf{w}_{u}^{\mathsf{T}} \mathbf{h}_{i} , c_{u,i}^{-1} ),
    \label{eq:wmf_nocontent}
\end{equation}
where $^{\mathsf{T}}$ denotes the matrix transpose and
\begin{equation}
    c_{u,i} = 1 + \alpha \log \left( 1 + \frac{y_{u,i}}{\epsilon} \right)
    \label{eq:confidence}
\end{equation}
is called the \emph{confidence}, with $\alpha = 2$ and $\epsilon = 10^{-6}$~\cite{Hu2008,Liang2015}. The model defined by~\eqref{eq:prior_w_h} and~\eqref{eq:wmf_nocontent} is estimated in a maximum a posteriori sense. Then, recommendation can be done based on the predicted ratings $\hat{r}_{u,i}=\mathbf{w}_u^{\mathsf{T}}\mathbf{h}_i$. However, this only applies to songs for which some listening history is available, hence facing the cold-start problem.

\subsection{Content-aware WMF}
\label{sec:model_content}

In order to incorporate content information in WMF, the authors in~\cite{Liang2015} propose to modify the prior on the item attributes matrix~\eqref{eq:prior_w_h}, which rewrites:
\begin{equation}
    \mathbf{h}_{i} \sim \mathcal{N}( \phi(\mathbf{z}_i),  \lambda_H ^{-1} \mathbf{I}_K ),
    \label{eq:wmfcontent_hi}
\end{equation}
where $\mathbf{z}_i \in \mathbb{R}^{L}$ is a content feature vector of dimension $L$, and $\phi$ is a mapping between this vector and the item attributes vector $\mathbf{h}_{i}$. The authors consider a linear mapping: $\phi(\mathbf{z}_i) = \mathbf{B} \mathbf{z}_i$ with $\mathbf{B} \in \mathbb{R}^{K \times L}$. Estimating the parameters with maximum a posteriori results in the following optimization problem:
\begin{multline}
 \min_{ \mathbf{W}, \mathbf{H}, \mathbf{B}} \sum_{u,i} c_{u,i} ( r_{u,i} - \mathbf{w}_u^{\mathsf{T}}\mathbf{h}_i )^2 \\+ \lambda_W \sum_u || \mathbf{w}_u ||^2 + \lambda_H \sum_i || \mathbf{h}_i - \mathbf{B} \mathbf{z}_i ||^2,
    \label{eq:min_wmf}
\end{multline}
where $||.||$ denotes the Euclidean norm. Canceling the gradient of the loss in~\eqref{eq:min_wmf} with respect to each parameter iteratively yields the following update rules:
\begin{align}
    \mathbf{w}_u &= (\mathbf{H} \text{diag}(\mathbf{c}_u) \mathbf{H}^{\mathsf{T}} + \lambda_W  \mathbf{I}_K )^{-1} \mathbf{H} \text{diag}(\mathbf{c}_u) \mathbf{r}_u, \\
    \mathbf{h}_i &= (\mathbf{W} \text{diag}(\mathbf{c}_i) \mathbf{W}^{\mathsf{T}} + \lambda_H  \mathbf{I}_K )^{-1} (\mathbf{W} \text{diag}(\mathbf{c}_i) \mathbf{r}_i + \lambda_H \mathbf{B} \mathbf{z}_i), \\
    \mathbf{B} &= \mathbf{H} \mathbf{Z}^{\mathsf{T}} (\mathbf{Z} \mathbf{Z}^{\mathsf{T}} + \lambda_B \mathbf{I}_L)^{-1}, 
\end{align}
where $\mathbf{r}_u = [r_{u,1},...,r_{u,I}]^{\mathsf{T}}$ and $\mathbf{r}_i = [r_{1,i},...,r_{U,i}]^{\mathsf{T}}$ (and similarly for $\mathbf{c}_u$ and $\mathbf{c}_i$), $\mathbf{Z} = [\mathbf{z}_1,...,\mathbf{z}_I]$, and $\lambda_B$ is a small offset added to the diagonal of $\mathbf{Z} \mathbf{Z}^{\mathsf{T}}$ in order to avoid numerical problems when inverting it~\cite{Liang2015}.

The advantage of this approach is two-fold. First, the content features serve as a regularizer for the collaborative filtering model parameters, and promotes a more meaningful representation learning. Second, this model can be used for solving the cold-start problem. Indeed, for a new item for which there is no user-item interaction, one can still predict a rating using $\hat{r}_{u,i}=\mathbf{w}_u^{\mathsf{T}} \mathbf{B} \mathbf{z}_i$, from which recommendation can be performed.

\section{Content features}
\label{sec:content}

\begin{figure}[t]
    \centering
    \includegraphics[width=.95\columnwidth]{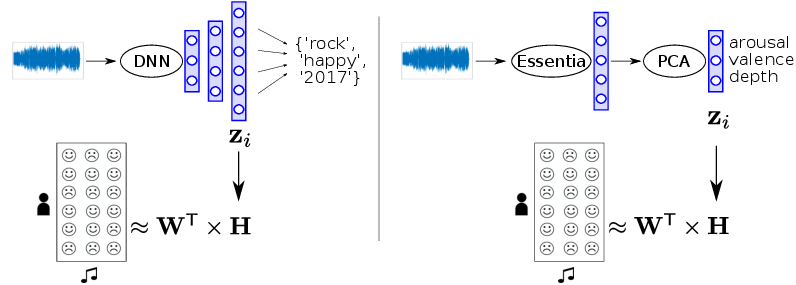}
    \caption{Illustration of the content-aware collaborative filtering: content features can be extracted using an auto-tagging DNN (left, as in~\cite{Liang2015}) or using the AVD model (right, proposed).}
    \label{fig:musrec_avd}
\end{figure}

In this section, we briefly describe the deep features used in~\cite{Liang2015}. We then motivate the use of the AVD model which we present subsequently. These two approaches are illustrated in Figure~\ref{fig:musrec_avd}.

\subsection{Deep features}
\label{sec:content_deep}

In~\cite{Liang2015}, the authors use as content features a latent representation that is extracted from an auto-tagging DNN, which is a system that aims at predicting music-related labels (such as genre) from music audio~\cite{Pons2018}. A set of $1028$ vector-quantized timbre features is calculated for each song, and fed as input to a DNN which predicts a set of tags associated with each song among $581$ possible tags. The proposed DNN consists of 3 feed-forward layers which serve as feature extractor, plus a logistic regression layer which serves as classifier. Once the tagging network is trained in a supervised fashion, the authors use the output of the last hidden layer as content features $\mathbf{z}_i$ of dimension $L=1200$.

This approach allows to alleviate the cold-start problem, but it suffers from several drawbacks. First, the quality of the content features is highly dependent on the performance of the auto-tagging network. While these methods have significantly improved in the recent years~\cite{Pons2018}, they are still limited by the noisy nature of the tags, which are user-annotated. Besides, this approach remains computationally demanding and the high dimensionality of the features suggests that it might be strongly redundant. Finally, the extracted content features are not guaranteed to characterize musical preference, and therefore lack a clear musical meaning, which we propose to overcome in the following.

\subsection{The AVD model}
\label{sec:content_avd}

Studies in the field of psychology of music have been conducted in order to design a model of musical preference. Early research has pointed out that the preference for musical genres can be described in terms of five factors, termed \emph{mellow}, \emph{unpretentious}, \emph{sophisticated}, \emph{intense}, and \emph{contemporary} (MUSIC)~\cite{Rentfrow2011}. The MUSIC model exhibited high correlation with genre,
but later studies~\cite{Greenberg2016,Fricke2018,Fricke2019} focused on more general musical preference attributes (i.e., non-genre specific). These studies outlined that musical preference can be conceptualized using three factors:
\begin{itemize}
    \item \emph{Arousal}, which describes how energetic and intense a music piece is (e.g., rousing or calm);
    \item \emph{Valence}, which is related to the perceived emotional content of a song (e.g., happy or sad);
    \item \emph{Depth}, which is a general indicator of level of sophistication (e.g., complicated or simple).
\end{itemize}
This AVD model has been shown to correlate well with high-level descriptors of music, whether those are expert-annotated~\cite{Fricke2018} or automatically computed~\cite{Fricke2019}. Since the AVD factors are a compact representation that is meaningful from a musical preference perspective, we claim that they are appropriate content features to incorporate in the collaborative filtering model presented in Section~\ref{sec:model_content}.

To compute the AVD factors, we follow the procedure used in~\cite{Fricke2019}. For each song, musical features are computed using the Essentia toolbox~\cite{Bogdanov2013}, a collection of a variety of music information retrieval algorithms. In a nutshell, Essentia takes as input the raw audio waveform and outputs a number of features, both low-level (e.g., spectral and temporal descriptors) and high-level (e.g., descriptors such as ``happy", ``sad" or ``danceable"). We retain the $16$ high-level features listed in Table~\ref{tab:avd_loadings}, which we scale to have zero mean and unit variance. We then perform a principal component analysis (PCA) with $3$ factors and oblimin rotation, which allows the factors to be non-orthogonal for better interpretability~\cite{Jolliffe2002}. The resulting principal components are the AVD factors, which we use as content features $\mathbf{z}_i$ of dimension $L=3$.

Note that we also considered a larger set of Essentia features comprising both low- and high-level descriptors ($84$ in total). This yields principal components with overall less correlation with high-level features. This suggests that the AVD model mostly captures the information contained in high-level features, and that adding some extra low-level information hampers the easiness to interpret these factors. Finally, this approach did not yield any significant improvement in terms of recommendation, thus we do not report hereafter the detailed results obtained in this setting.

\section{Experiments}
\label{sec:exp}

In this section, we present the experiments conducted to assess the potential of the AVD model for music recommendation. In the spirit of reproducible research, we provide the code related to these experiments,\footnote{\url{https://github.com/magronp/mus-reco-avd}} as well as the Essentia features data.\footnote{\url{https://zenodo.org/record/3860557#.Xs-b7R9fg5l}}

\subsection{Protocol}
\label{sec:exp_protocol}

\paragraph*{Dataset.}
We use the Taste Profile dataset which is part of the Million Song Dataset~\cite{BertinMahieux2011}. It provides listening counts of $1$ million users and $380,000$ songs. We only keep the songs whose Essentia (and consequently AVD) features can be calculated, which corresponds to a total of $204,316$ songs~\cite{Fricke2019}. 
The playcount data is binarized by retaining values of five or higher as implicit feedback~\cite{Tran2019}. As in~\cite{Liang2015,Gouvert2018}, in order to keep the computational burden low, we retain the top songs and users (sorted by playcounts) and we remove inactive users and items (that is, we only keep users who listened to at least 20 songs, and songs which have been listened to by at least 50 users). The resulting dataset comprises $9,132$ users and $7,674$ songs, for a total of $247,414$ ($0.35$ $\%$) non-zero playcounts.

\noindent We use $95$~$\%$ of the songs for training the WMF model, which corresponds to an \emph{in-matrix} recommendation task. That is, the corresponding playcounts are split into 70~\%, 20~\% and 10~\% for training, validation and in-matrix testing, respectively. The playcounts corresponding to the remaining $5$~$\%$ of the songs are used for an \emph{out-of-matrix} recommendation task. This corresponds to the cold-start scenario, since the WMF model has not been trained on these songs.

\paragraph*{Methods.}
We compare the content-free WMF (used as a baseline for in-matrix recommendation) to its content-aware counterpart. As for content features, we use the AVD factors estimated according to the protocol described in Section~\ref{sec:content_avd}, as well as the Essentia features before PCA. Even though the deep features presented in Section~\ref{sec:content_deep} would constitute an interesting comparison reference, we were not able to reproduce the results from~\cite{Liang2015} as we faced some technical issues when re-implementing the auto-tagging network. Following  previous studies~\cite{Liang2015}, WMF is run with $20$ iterations and rank $K=50$. The hyperparameters $\lambda_W$ and $\lambda_H$ are tuned on the validation set using the normalized discounted cumulative gain (NDCG, see below), and $\lambda_B= 10^{-2}$.

\noindent Since content-free WMF cannot perform an out-of-matrix recommendation task, we implemented a pure content-based method as a baseline for cold-start recommendation based on~\cite{Soleymani2015}. This method consists in computing a mean AVD preference vector for each user from the training set, and then performing recommendations based on similarities between this mean AVD preference vector and the AVD factors extracted from novel songs. Note that the original method in~\cite{Soleymani2015} used the MUSIC factors~\cite{Rentfrow2011}, but we use here the AVD factors for a fair comparison with our proposed content-aware collaborative filtering technique.

\paragraph*{Evaluation metric.}
We use the NDCG~\cite{Wang2013} as a measure of the overall quality of the recommendation. For each user $u$, we compute a ranked list of items in the test set  based on the
predicted preference $\hat{\mathbf{r}}_u$. We then compute the relevance of this list with respect to the ground truth preferences (that is, the ratings that were left out for testing): $\text{rel}_{u,i} = 1$ if the item $i$ is in the listening history of user $u$ and $0$ otherwise. In order to favor recommendations that place the test items high in the list, we apply a discounted weight to the relevance, which yields the discounted cumulative gain (DCG), and from which its normalized version (ranging from 0 to 1) can be obtain:
\begin{equation}
    \text{DCG}_u = \sum_{i=1}^I \frac{\text{rel}_{u,i}}{ \log_2(i+1) } \text{, } \text{NDCG}_u = \frac{\text{DCG}_u}{ \text{IDCG}_u },
\end{equation}
where IDCG is the ideal DCG, which corresponds to the DCG of a perfectly ranked list. Finally, these scores are averaged over users to yield an overall recommendation performance.

\subsection{Features analysis}

First, we analyze the AVD factors computed according to the protocol presented in Section~\ref{sec:content_avd}. We report in Table~\ref{tab:avd_loadings} the correlation between the high-level Essentia features and the AVD factors. We observe that these correlations are consistent with the definition of these factors. Indeed, the arousal component positively correlates with features such as ``rousing" or ``aggressive" and negatively correlates with ``relaxed" or ``sad".  The valence factor has a high loading on features such as ``happy", ``fun" and ``danceable", while it is negatively correlated with ``aggressive" and ``intense". In order to subjectively assess the validity of these factors, we examine the songs with maximum and minimum AVD values (the interested reader can use the provided code to replicate and extend this analysis). We observe that maximum arousal songs belong to the punk, rock or metal genres, and exhibit relatively high aggressiveness and energy, while low-valence pieces are mostly tracks with a dark and/or sad atmosphere and lyrical content. These subjective assessments are consistent with the correlations between high-level Essentia features and the AVD factors highlighted in Table~\ref{tab:avd_loadings}.

The meaning of the depth factor is less clear: it highly correlates with ``electronic" and its lowest loading is on the ``tonal" feature, which suggests that this factor may characterize contemporary and intricate pieces, which are indeed encountered among the top-depth songs. However, its high correlation with the ``danceable" feature might appear as contradictory with this description. Note that this ambiguity of the depth factor, as well as these general trends, are overall consistent with the findings of previous studies~\cite{Fricke2018,Fricke2019}.

\begin{table}[t]
\centering
    \caption{Correlations between the high-level Essentia features and the AVD factors.}
    \label{tab:avd_loadings}
    \begin{tabular}{lccc} 
    \toprule
    & Arousal & Valence & Depth  \\
    \midrule
Mirex clusters & & & \\
\hspace{0.5em} 1 (Rousing, Passionate)    & \textbf{0.56} &-0.11 &-0.12 \\  
\hspace{0.5em} 2 (Fun, Cheerful)          &-0.03 & \textbf{0.78} & 0.01 \\  
\hspace{0.5em} 4 (Humorous, Witty)        &-0.05 & \textbf{0.63} & 0.12 \\  
\hspace{0.5em} 5 (Aggressive, Intense)    & 0.34 &\textbf{-0.55} & 0.30 \\  
Mood-related & & & \\
\hspace{0.5em} Aggressive                 & \textbf{0.63} &-0.48 &-0.02 \\  
\hspace{0.5em} Happy                      & \textbf{0.52} & 0.37 &-0.36 \\  
\hspace{0.5em} Party                      & \textbf{0.69} & 0.05 & 0.40 \\  
\hspace{0.5em} Relaxed                    &\textbf{-0.84} & 0.01 & 0.14 \\  
\hspace{0.5em} Sad                        &\textbf{-0.80} & 0.07 &-0.21 \\  
Sound-related & & & \\
\hspace{0.5em} Acoustic                   &\textbf{-0.78} & 0.04 &-0.25 \\  
\hspace{0.5em} Average loudness           & \textbf{0.59} & 0.14 &-0.07 \\  
\hspace{0.5em} Danceable                  & 0.23 & 0.42 & \textbf{0.52} \\  
\hspace{0.5em} Dissonance                 & \textbf{0.86} &-0.03 &-0.04 \\  
\hspace{0.5em} Dynamic complexity         &\textbf{-0.57} & 0.07 & 0.21 \\  
\hspace{0.5em} Electronic                 & 0.08 &-0.01 & \textbf{0.74} \\  
\hspace{0.5em} Instrumental               &-0.35 &-0.06 & 0.23  \\  
\hspace{0.5em} Tonal                      & 0.04 & 0.15 &\textbf{-0.60} \\  
    \bottomrule
    \end{tabular}
\end{table}

\subsection{Recommendation results}
\label{sec:exp_recom}

Table~\ref{tab:ndcg} presents the results in terms of NDCG for in- and out-of-matrix recommendation. For in-matrix recommendation, the performance of the methods are similar. A slight improvement is obtained when incorporating some content information within WMF in the form of Essentia features, but this improvement is no longer observed when using the AVD model instead. This framework is known to work quite well on those datasets~\cite{Liang2015}, thus this collaborative filtering method do not benefit greatly from adding extra content information for recommending items which already have some listening history.

The advantage of content-aware WMF appears more clearly for out-of-matrix recommendation, where the usage of the proposed features alleviates the cold-start problem. Using the set of Essentia features slightly outperforms using the AVD factors. Overall, both content-aware WMF methods outpeform the pure content baseline. This reveals the appropriateness of the AVD model for addressing the cold-start problem, and it particularly outlines its relevance when combined with user-iterm interaction data in a collaborative filtering-based framework.

\begin{table}[t]
    \centering
	\caption{Recommendation performance expressed with the NDCG averaged over users (higher is better).}
	\label{tab:ndcg}
	\begin{tabular}{lcc}
    \toprule
		 & In-matrix & Out-of-matrix \\
		 \midrule
		 Content-free WMF     & $0.35$ & $-$  \\
		 Pure content     & $-$ & $0.19$  \\
		 \hline
		 Content-aware WMF   & &   \\
		\hspace{0.5em} Essentia   & $0.36$ &  $0.22$ \\
		\hspace{0.5em} AVD        & $0.35$ &  $0.21$ \\
       \bottomrule
	\end{tabular}
\end{table}

Finally, unlike the approach in~\cite{Liang2015} which used deep features, our approach is light (it does not require to train an auto-tagging DNN) and relies on content features with a clear musical meaning.

\section{Conclusion}
\label{sec:conclusion}

In this paper, we proposed to leverage the AVD model of musical preference for music recommendation, which was shown to correlate well with high-level musical properties. We extracted the AVD factors from audio excerpts and integrated them into a content-aware collaborative filtering method. This approach has shown its potential for music recommendation, notably for addressing the cold-start problem in a computationally very light framework. 

Future work will focus on combining this model with other types of content (such as tags or lyrics) in order to fully exploit the available data for recommendation. Finally, we will also explore alternative mappings between content features and item attributes, such as non-linear and/or deep.

\section{Acknowledgments}

This work is supported by the European Research Council (ERC FACTORY-CoG-6681839). We thank K. R. Fricke, D. M. Greenberg and P. J. Rentflow for fruitful discussion about the AVD model and for providing the data used in this work. We also thank Olivier Gouvert for his insightful comments and for proof-reading the paper.

\bibliographystyle{IEEEbib}
\bibliography{references}

\end{document}